\documentclass[sigconf]{acmart}

\usepackage{xspace,framed}
\usepackage{xcolor}
\usepackage{amssymb,amsmath,amsfonts} 
\usepackage{mathtools}
\usepackage{algorithm}
\usepackage{tabularx}
\usepackage{listings}
\usepackage{afterpage}
\usepackage{graphicx}
\usepackage{color}
\usepackage{multirow}
\usepackage[latin1]{inputenc}
\usepackage{amsmath}
\usepackage{amssymb}
\usepackage{marvosym}
\usepackage{caption}
\usepackage{pgf}
\usepackage{tikz}
\usepackage{subfig}
\usepackage[american]{babel}
\usetikzlibrary{arrows,automata,positioning}
\usepackage{epstopdf}
\usepackage{afterpage}
\newtheorem{mydef}{Definition}

\DeclareGraphicsExtensions{.pdf,.png,.jpg}

\setcopyright{acmcopyright}
\acmDOI{}
\acmISBN{}
\acmConference[]{}{}{} 
\acmYear{}
\copyrightyear{}
\acmPrice{}

\begin{document}

\newcommand\tool{{\sf DSValidator}\xspace}
\newcommand{\fwl}[1]{\mathcal{FWL}[#1]}
\newcommand{\roots}[1]{\mathcal{R}[#1]}
\newcommand{\reviews}[1]{\textbf{\textcolor{blue}{#1}}}

\title{DSValidator\:: An Automated Counterexample Reproducibility \\ Tool for Digital Systems}

\author{Lennon Chaves, Iury Bessa}
\affiliation{%
  \institution{Federal University of Amazonas}
  \country{Manaus, Brazil}
}
\email{lennonchaves@ufam.edu.br, iurybessa@ufam.edu.br}

\author{Lucas Cordeiro, Daniel Kroening}
\affiliation{%
  \institution{University of Oxford}
  \country{Oxford, United Kingdom}
}
\email{lucas.cordeiro@cs.ox.ac.uk, kroening@cs.ox.ac.uk}

\begin{abstract}
An automated counterexample reproducibility tool based on 
MATLAB is presented, called \tool, with the goal of reproducing 
counterexamples that refute specific properties related to digital systems. 
We exploit counterexamples generated by the Digital System Verifier (DSVerifier), 
which is a model checking tool based on satisfiability modulo theories for digital systems. 
\tool reproduces the execution of a digital system, relating its input with the counterexample, 
in order to establish trust in a verification result. We show that \tool 
can validate a set of intricate counterexamples for digital controllers used 
in a real quadrotor attitude system within seconds and also expose incorrect verification results in
DSVerifier. The resulting toolbox leverages the potential of combining different verification tools for validating 
digital systems via an exchangeable counterexample format.
\end{abstract}


\maketitle

%-----------------------------------
\section{Introduction}
%-----------------------------------
Digital systems ({\it e.g.}, filters and controllers) consist of a mathematical operator 
that maps one signal into another signal using a fixed set of operations~\cite{diniz}; 
they are used in several applications due to advantages over the analog 
counterparts, such as reliability, flexibility, and cost. However, there are disadvantages 
in the use of digital systems; since they are typically implemented in microprocessors, 
errors might be introduced due to quantizations and round-offs. The hardware choice, 
realizations ({\it e.g.}, delta and direct forms), and implementation features ({\it e.g.}, 
number of bits, fixed-point arithmetic) impact the digital system precision and 
performance~\cite{daes20161}.

To detect errors in a digital system implementation, considering finite word-length (FWL) 
effects~\cite{fadali09,istepanianbook01}, a model checking procedure based on satisfiability 
modulo theories (SMT) has been proposed in the literature, named as Digital System Verifier 
(DSVerifier)~\cite{dsverifier}, which verifies digital filters and controllers represented 
by transfer functions and state-space equations~\cite{daes20161,Bessa16,monteiro2016,Abreu2016}. 
DSVerifier checks properties related to overflow, limit cycle, stability, and minimum-phase in 
different digital system realizations and numerical formats. 

Currently, there are several toolboxes in MATLAB to support digital system design~\cite{matlab-toolbox}. 
For instance, the fixed-point designer toolbox provides data-types and tools for developing fixed-point digital systems. 
There are also other toolboxes implemented with different objectives, {\it e.g.}, optimization, control system, 
and digital signal processing~\cite{matlab-toolbox}. However, there is no toolbox to reproduce counterexamples 
in digital systems generated by verifiers, {\it i.e.}, automatically reproduce a sequence of states 
that refutes a specific property in order to establish trust in a verification result.

There are verifiers to validate counterexamples using the witness validation approach, 
which reproduces the verification results by checking a given counterexample based 
on the graphml format~\cite{dirk2015}. For instance, CPAchecker~\cite{cpachecker} and 
Ultimate Automizer~\cite{ultimate}  employ the error-witness-driven program analysis technique 
to avoid false alarms produced by verifiers, {\it i.e.}, given a  witness for a problematic program path, 
they re-verify that the witness indeed violates the specification. However, those tools are currently 
unable to support the validation of systems that require fixed-point arithmetic, and consequently 
disregard FWL effects, which is needed to successfully validate implementation-level properties of typical digital systems.

\textit{Contributions}. This paper describes and evaluates a novel MATLAB toolbox called \tool to 
automatically check whether a counterexample given by a verifier is reproducible. 
We propose a particular format to represent counterexamples, which can be used by other verifiers 
and MATLAB toolboxes. Here, a counterexample provides assignments to the 
digital system's variables, which can be extracted to reproduce a given violation. This counterexample 
allows us to reproduce the failed property, providing concrete, lower-level details 
needed to simulate the digital system in MATLAB. \tool can validate counterexamples related to 
overflow, limit-cycle, stability, and minimum-phase. We show that \tool is able to reproduce a set of 
intricate counterexamples for digital controllers used in a real quadrotor attitude control system within seconds. 
\tool can also expose incorrect verification results in DSVerifier due to wrong computations of the 
system output.

\textit{Availability of Data and Tools}. Our experiments are based on a set of publicly available benchmarks. 
All tools, benchmarks, videos, and results of our evaluation are available on a supplementary 
web page \texttt{http://dsverifier.org/}. In particular, \tool source code is available in a public repository 
located at \url{https://github.com/ssvlab/dsverifier/tree/master/toolbox-dsvalidator}.

%-----------------------------------
\section{DSValidator Digital System Reproducibility Engine}
\label{sec:engine}
%-----------------------------------

\tool is able to simulate digital controllers and filters considering implementation 
features ({\it e.g.}, FWL effects and realizations) by taking into account a given counterexample 
provided by a verifier.

%-----------------------------------
\subsection{Digital System Representation}
\label{ssec:representation}
%-----------------------------------

\tool supports digital systems (digital controllers and filters), represented by transfer functions, 
{\it i.e.}, frequency domain equations that are able to represent input-to-output relations in a digital system. 
The following expression presents the general form of a digital system transfer function:
\begin{equation}
\label{eq:transferfunction}
H(z)=\frac{B(z)}{A(z)}=\frac{b_{0}+b_{1}z^{-1}+...+b_{M}z^{-M}}{a_{0}+a_{1}z^{-1}+...+a_{N}z^{-N}},
\end{equation}
where $z^{-1}$ is called backward-shift operator; $A(z)$ and $B(z)$ are the denominator and numerator polynomials; 
and $N$ and $M$ represent the denominator and numerator polynomials order, respectively. 
Another representation is the difference equation, which can be described as

\begin{equation}
y(n)=-\sum_{k=1}^{N} a_{k}y(n-k)+\sum_{k=0}^{M}b_{k}x(n-k).
\label{eq:diffeq}
\end{equation}

Eq.~\eqref{eq:diffeq} allows \tool to compute the system output $y(n)$ at the $n$-th instant ({\it i.e.}, 
at time $t=n\cdot T$, where $T$ is the system sample time) using values of the past outputs and 
the present and past inputs ({\it i.e.}, $x(n)$).

%-----------------------------------
\subsection{Counterexample Reproducibility for the Digital System Properties}
\label{ssec:simulation}
%-----------------------------------

%-----------------------------------
\subsubsection{Stability and Minimum-phase.}
\label{ssec:stability}
%-----------------------------------

A digital system is stable \textit{iff} all of its poles are inside the $z$-plane unitary circle; 
poles must have the modulus less than one. Minimum-phase is also a desirable property 
for digital systems. A digital system is a minimum-phase system \textit{iff} all of its zeros 
are inside the $z$-plane unitary circle. The counterexample reproducibility for both 
minimum-phase and stability does not require \tool to compute output and states 
since polynomial analysis is performed, but FWL effects over the coefficients of 
Eq.~\eqref{eq:transferfunction} must be computed. 

%\begin{mydef}
%\label{counterexample}
%\textbf{(Counterexample)}
%A counterexample for a property $\phi$ represents a sequence of states $s_{0}, s_{1},\ldots, s_{k}$ with $s_{0} \in S_{0}$ (initial state), $s_{k} \in S_k$ (bad state) that refutes $\phi$. 
%\end{mydef}

\begin{mydef}
\label{fwldef}
\textbf{(Finite Word-Length)} 
$\fwl{\cdot}:\mathbb{R}^{N+M+2}\rightarrow Q[\mathbb{R}^{N+M+2}]$ function 
applies FWL effects to a polynomial vector representation, where 
$Q[\mathbb{R}]$ represents the quantized set of representable real numbers in 
the chosen implementation format.
\end{mydef}

\begin{mydef}
\label{rootdef}
\textbf{(Roots of a Polynomial)} 
$\roots{\cdot}:\mathbb{R}^{N+M+2}\rightarrow \mathfrak{S}$ function computes 
the set of roots of a polynomial, and $\mathfrak{S}$ is a family of sets. The 
poles of Eq.~\eqref{eq:transferfunction} is computed by $\roots{A(z)}$, and the 
zeros are computed by $\roots{B(z)}$.
\end{mydef}

\begin{mydef}
\label{stabilitydef}
\textbf{(Stability Reproducibility)} 
\tool computes the $\fwl{A(z)}$ roots for stability reproduction. If any root has modulus equal or greater than one, then the system is unstable; otherwise, it is stable. 
\end{mydef}

\begin{mydef}
\label{minimumphasedef}
\textbf{(Minimum-phase Reproducibility)}
\tool computes the $\fwl{B(z)}$ roots for minimum-phase reproduction. If any root has modulus equal or greater than one, then the system is non minimum-phase; otherwise, it is minimum-phase.  
\end{mydef}

\subsubsection{Overflow}
\label{ssec:overflow}
%-----------------------------------

When an operation result exceeds the limited range of the processor's word-length, overflow might occur in the output of the digital system realization, resulting in undesirable nonlinearities in the output. In order to reproduce an overflow counterexample, the output sequence must be computed for the given input sequence; the counterexample should contain an input sequence $x(n)$ that leads the digital system to overflow. \tool reads the counterexample provided by a given verifier, and then computes FWL effects over the coefficients, {\it i.e.}, \tool computes $\fwl{A(z)}$ and $\fwl{B(z)}$ (cf. Definition~\ref{fwldef}). 

After that, \tool checks the word-length representation limits, considering $n$-integer bits and $l$-fractional bits. The maximum representable value is computed as $2^{n-1}-2^{-l}$ and the minimum representable value is computed as $-2^{n-1}$. Then, Eq.~\eqref{eq:diffeq} is iteratively unrolled for a given realization form, considering the input $x(n)$ (from the counterexample) to produce the output $y(n)$. 

\begin{mydef}
\label{realizationformdef}
\textbf{(Realization Form)}
A realization form represents a template to implement a given digital system in software by using directly the coefficients of Eq.~\eqref{eq:transferfunction} in its implementation~\cite{daes20161}.
\end{mydef}

\begin{mydef}
\label{overflowdef}
\textbf{(Overflow reproducibility)}
\tool checks whether each system's output is inside the word-length representation limits; the output does not lead to overflow if $-2^{n-1}<y(n)<2^{n-1}-2^{-l}$ is inside the word-length limits. A detected overflow violation must be similar to the counterexample indicated by the verifier; otherwise, the counterexample is not reproducible. 
\end{mydef}

\subsubsection{Limit Cycle Oscillation (LCO)}
\label{ssec:limit-cycle}
%-----------------------------------

Limit cycle is defined by the presence of oscillations in the output, even when the input sequence is constant. LCO can be classified as granular or overflow. 

\begin{mydef}
\label{granularlcodef}
\textbf{(Granular LCO)}
Granular limit cycles are autonomous oscillations due to round-offs in the least significant bits~\cite{diniz}. 
\end{mydef}

\begin{mydef}
\label{overflowlcodef}
\textbf{(Overflow LCO)}
Overflow limit cycles appear when an operation results in overflow using the wrap-around mode~\cite{diniz}. 
\end{mydef}

To reproduce LCO counterexamples, constant inputs and initial states are used 
as test signals in \tool to compute the output sequence $y(n)$, considering a given 
realization form (cf. Definition~\ref{realizationformdef}). First, \tool obtains FWL 
effects on the numerator and denominator coefficients (cf. Definition~\ref{fwldef}). 
The constant input, initial states, and realization form are provided by a given 
counterexample and employed to compute $y(n)$ based on the fixed-point arithmetic 
and also to simulate the respective digital system in MATLAB. 

\begin{mydef}
\label{lcodef}
\textbf{(LCO reproducibility)}
If the system's output $y(n)$ provided by \tool leads to oscillations in the output with the same characteristics ({\it i.e.}, amplitude and period) from that indicated by the verifier, then the LCO counterexample is reproducible; otherwise, the verifier presents an error. 
\end{mydef}

In order to confirm the LCO absence, the algorithm proposed by Bauer~\cite{bauer,premaratne} was implemented in \tool. The aim of that algorithm is to exhaustively search for the absence of limit cycle; it is applicable to all direct form realizations, besides being independent on the quantization and system order. Therefore, Bauer's method decides about the asymptotic stability of (linearly stable) digital systems, by employing an exhaustive search method. If it detects that a digital system is asymptotic stable, then the latter is limit cycle free; otherwise, it is susceptible to overflow or granular LCO. 
%A LCO example can be seen in the following digital system:
%%
%\begin{equation}
%\label{eq:eq-limit}
%H(z)=\frac{2002 -4000z^{-1} +1998z^{-2}}{1.0-1.0z^{-2}}.
%\end{equation}
%%
%The FWL effects in Eq.~\eqref{eq:eq-limit} using 12 bits for the integer and 6 bits for the fractional part lead to a digital system with the same coefficients. However, the digital system represented by Eq.~\eqref{eq:eq-limit} exhibits overflow LCO (cf. Definition~\ref{granularlcodef}), as shown in Table~\ref{table:limit-example}, considering $x(n)$ for a given realization form (cf. Definition~\ref{lcodef}).
%%(Direct Form II)
%%
%\begin{table}[H]
%\setlength{\tabcolsep}{2pt}
%\renewcommand\arraystretch{1.18}
%\begin{center} {
%\begin{tabular}
%{ | c | c | c | c | c | c | c | c | c |}
%\hline
%n & $1$ & $2$ & $3$ & $4$ & $5$ & $6$ & $7$ & $8$ \\
%\hline
%$x(n)$ & $-0.0781$ & $-0.0781$ & $-0.0781$ & $-0.0781$ & $-0.0781$ & $-0.0781$  & $-0.0781$  & $-0.0781$ \\
%\hline
%$y(n)$ & $-0.1284$  & $-2.0480$ & $-0.1284$  & $-2.0480$ & $-0.1284$  & $-2.0480$ & $-0.1284$  & $-2.0480$ \\
%\hline
%\end{tabular} }
%\end{center}
%\caption{A limit cycle example for the digital system represented in Eq.~\eqref{eq:eq-limit}.}
%\label{table:limit-example}
%\end{table}

%-----------------------------------
\section{Automated Counterexample Reproducibility for Digital Systems}
\label{Methodology}
%-----------------------------------
%
%---------------------------------------------------------
\subsection{Proposed Counterexample Format}
\label{sec:counterexample-format}
%---------------------------------------------------------

\tool exploits counterexamples provided by verifiers~\cite{dsverifier,cbmc,esbmc}; 
if there is a property violation, then the verifier provides a counterexample, which contains 
inputs and initial states that lead the digital system to a failure state. 
Fig.~\ref{fig-counterexample-dsv} shows an example of the present counterexample 
format related to an overflow LCO violation for the digital system represented by Eq.~\eqref{control_1}:
\begin{equation}
H(z)=\frac{2002-4000z^{-1}+1998z^{-2}}{1-z^{-2}}.
\label{control_1}
\end{equation}
\begin{figure}[ht]
\begin{lstlisting}[xleftmargin=.01\textwidth,frame=single,]
Property = LIMIT_CYCLE
Numerator  = { 2002,  -4000,  1998 }
Denominator  = { 1,  0,  -1 }
X_Size = 10
Sample_Time = 0.001
Implementation = <13,3>
Numerator (fixed-point) = { 2002, -4000, 1998 }
Denominator (fixed-point) = { 1, 0, -1 }
Realization = DFI
Dynamical_Range = { -1, 1 }
Initial_States = { -0.875, 0, -1 }
Inputs = { 0.5, 0.5, 0.5, 0.5, 0.5, 0.5, 0.5, 0.5, 0.5, 0.5}
Outputs = { 0, -1, 0, -1, 0, -1, 0, -1, 0, -1}
\end{lstlisting}
\caption{Proposed Counterexample Format Example.}
\centering\scriptsize
\label{fig-counterexample-dsv}
\end{figure}

The proposed counterexample format shown in Fig.~\ref{fig-counterexample-dsv} describes 
the violated property (represented by a \textit{string}), transfer function numerator and 
denominator (represented by \textit{fixed-point numbers}), bound (represented by an \textit{integer}), 
sample time (represented by a \textit{fixed-point number}), implementation aspects 
(integer and fractional bits represented by an \textit{integer}), realization form 
(represented by a \textit{string}), dynamical range (represented by an \textit{integer}), 
initial states, inputs, and outputs (which are represented by \textit{fixed-point numbers}). 
In particular, the counterexample provides the needed data to reproduce a given property 
violation  via simulation in MATLAB. 

\tool considers ``.out'' files to extract the counterexample and to transform them 
into MATLAB variables; those ``.out'' files are generated by the verifier after the digital 
system verification is performed. 
%Particularly, \tool employs the attributes provided by the current counterexample format, and then computes the outputs in order to reproduce the violation in MATLAB. 
Currently, \tool is able to validate the minimum-phase, overflow, stability, and limit-cycle 
properties for a digital system that is represented by a transfer function. 
Additionally, \tool is able to employ $6$ direct and delta realization forms for digital systems: 
direct form I (DFI), direct form II (DFII), transposed direct form II (TDFII), delta direct form I (DDFI), 
delta direct form II (DDFII), and transposed delta direct form II (TDDFII)~\cite{daes20161}.

\subsection{Automated Counterexample Validation}
\label{Stepts-of-Automatic-Validation}
%-----------------------------------

There are five steps to automatically perform the automated counterexample validation in \tool.
%
%\begin{figure}[H]
%\centering
%\includegraphics[width=0.5\textwidth]{validation_process.pdf}
%\caption{Automatic Counterexample Validation Process.}
%\label{fig_steps}
%\end{figure}
%
In step ($1$), \tool obtains the counterexample and then uses a shell script to extract the 
data related to the digital system, {\it i.e.}, property, transfer function numerator and denominator, 
fixed-point representation, \textit{k}-bound, sample time, implementation aspects, realization form, 
dynamical range, initial states, inputs, and outputs. In step ($2$), \tool converts all counterexample 
attributes into variables that can be manipulated in MATLAB. 
In step ($3$), \tool simulates the counterexample (violation) for the failed property, 
which is derived from the counterexample by providing concrete, lower-level details needed to 
simulate the digital system in MATLAB. In this specific step, all FWL effects are applied 
to the digital system, and computations to perform the outputs are produced, according to the 
realization form and property, as previously mentioned (cf. subsection~\ref{ssec:simulation}). 
In step ($4$), \tool compares the result between the output provided by the verifier and that simulated 
by MATLAB. Finally, in step ($5$), \tool stores the extracted counterexample 
in a \texttt{.MAT} file and then reports its reproducibility.

%-----------------------------------
\subsection{\textbf{DSValidator} Features}
\label{Toolbox-Features}
%-----------------------------------

\tool's features can be described as follows:\footnote{Functions implemented in \tool are described in the \href{http://dsverifier.org/matlab-toolbox/dsvalidator-documentation/}{Toolbox Documentation}.}

\begin{itemize}
\item{\textbf{Macro Functions:} functions to reproduce the validation steps ({\it e.g.}, parsing, simulation, comparison, and report).}
\item{\textbf{Validation Functions:} check and validate a violated property ({\it e.g.}, overflow, limit-cycle, stability, and minimum-phase).}
\item{\textbf{Realizations:} reproduce realizations forms to validate overflow and limit-cycle (for direct and delta forms).} 
\item{\textbf{Numerical Functions:} perform the quantization process; select rounding mode and overflow mode (wrap-around and saturate); fixed-point operations ({\it e.g.}, sum, subtraction, multiplication, division); and delta operator.}
\item{\textbf{Graphic Functions:} plot the graphical representation of overflow to show each output exceeding the supported word-length limits; limit-cycle to represent the system's output oscillations; and poles/zeros to show stability and minimum-phase with (or without) FWL effects inside a unitary circle.}
\end{itemize}

%-----------------------------------
\subsection{\textbf{DSValidator} Result}
\label{Counterexample-Features}
%-----------------------------------

\begin{figure}[ht]
\centering
\includegraphics[width=0.5\textwidth]{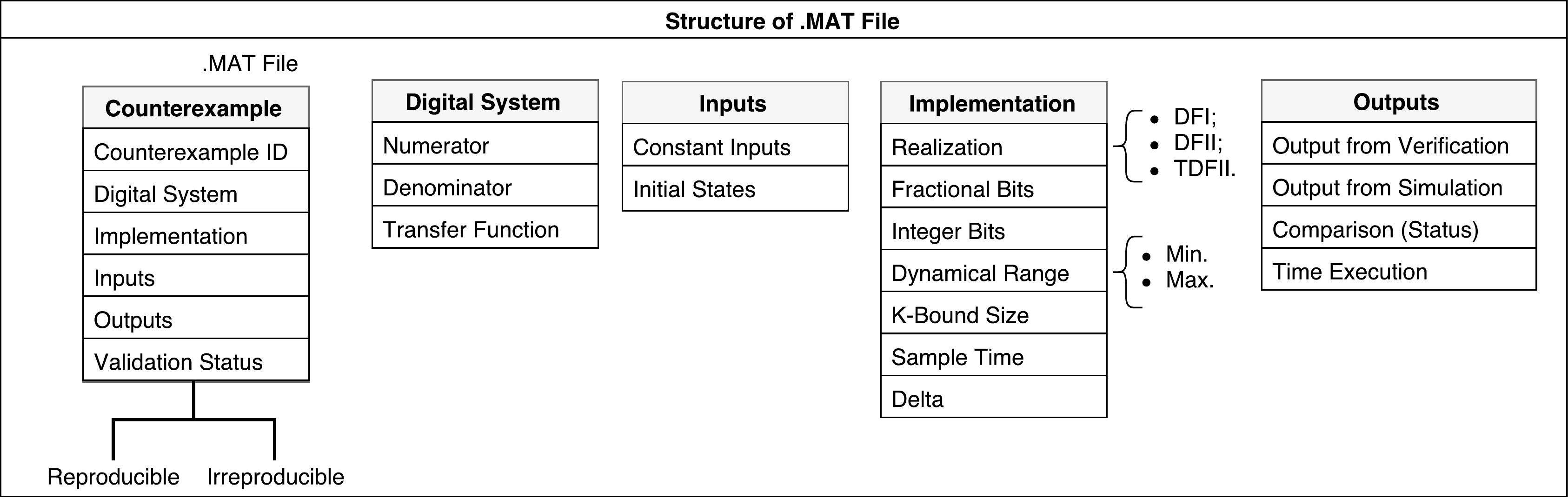}
\caption{Structure of the \texttt{.MAT} file for representing counterexamples.}
\label{fig2}
\end{figure}

The \tool result is structured with counterexample data composed by attributes and classes as shown in Fig.~\ref{fig2}.
The attributes are defined in the \texttt{.MAT} file with the following structure: \textit{counterexample} that represents the counterexample identification; \textit{digital system} that represents the numerator, denominator, and transfer function representation;  \textit{inputs} that represent the input vector and initial states; \textit{implementation} that represents the integer and fractional bits, dynamical ranges, delta operator, sample time, bound, and realization form; \textit{outputs} report the verification and simulation results, execution time in MATLAB, and comparison status, where it reports whether the counterexample is reproducible or not. Importantly, all execution times are actually CPU times, {\it i.e.}, only the elapsed time periods spent in the allocated CPUs, which is measured with the {\tt times} system call (POSIX system).

%-----------------------------------
\subsection{\textbf{DSValidator} Usage}
\label{Toolbox-Usage}
%-----------------------------------

\tool can be called via command line in MATLAB as:
\begin{center}
\texttt{validation(path, property, ovmode, rmode, filename)}
\end{center}
\noindent where \texttt{path} is the directory with all counterexamples; \texttt{property} 
is defined as: \textbf{``m''} for minimum phase; \textbf{``s''} for stability; \textbf{``o''} for overflow; 
and \textbf{``lc''} for limit cycle; \texttt{ovmode} represents the overflow mode: \textbf{``wrap''} 
for wrap-around mode (default) and \textbf{``saturate''} for saturation mode; \texttt{rmode} 
represents the rounding mode, which can be \textbf{``round''} (default) and \textbf{``floor''}; 
\texttt{filename} represents the \texttt{.MAT} filename, which is generated after the 
validation process; by default, the \texttt{.MAT} file is named as \texttt{digital\_system}.
After executing the \texttt{validation} command, \tool prints statistics about the 
counterexamples validation. Fig.~\ref{fig-valid-report} shows a report about the 
digital system represented in Eq.~\eqref{control_1} for realizations DFI, DFII, and TDFII.
%\tool validates each individual property as: \texttt{ReproduceOverflow}; \texttt{ReproduceLimitCycle}; \texttt{ReproduceMinimumPhase}; and \texttt{ReproduceStability}, with the counterexample path as input argument. 

\begin{figure}[ht]
\begin{lstlisting}[xleftmargin=.02\textwidth,frame=single,]
Running Automatic Validation...
Counterexamples (CE) Validation Report...
CE 1 time: 0.081929 status: reproducible
CE 2 time: 0.013996 status: reproducible
CE 3 time: 0.009488 status: reproducible
General Report:
Total Counterexamples Reproducible: 3
Total Counterexamples Irreproducible: 0
Total Counterexamples: 3
Total Execution Time: 0.10541
\end{lstlisting}
\caption{Counterexample Reproducibility Report.}
\label{fig-valid-report}
\end{figure}

%-----------------------------------
\section{Case Study: Controllers for UAVs}
\label{case-study}
%-----------------------------------
%-----------------------------------
\subsection{Benchmarks Description}
\label{benchmark-description}
%-----------------------------------

Here, we evaluated \tool for a set of $11$ digital controllers 
extracted from a real quadrotor unmanned aerial vehicle (UAV)~\cite{bouabdallah}. 
\tool has also been evaluated with other control system benchmarks 
and experimental results are available online.\footnote{\url{http://www.dsverifier.org/benchmarks}}
The experiments for the UAV controllers evaluate overflow, minimum-phase, stability, and limit-cycle 
in $33$ different numerical formats: $3$ for each digital controller, 
using $3$ different realizations forms ({\it i.e.}, DFI, DFII, and TDFII), 
which resulted in $99$ different verification tasks for each property 
($396$ verification tasks in total). The chosen number of bits, 
associated to each implementation, is based on the impulse response sum 
proposed by Carletta {\it et al.}~\cite{carletta}. 

%the methodology presented by Carletta {\it et al.}, which suggests a computation based on 

%All implementations and realizations used in the experiments are available online\footnote{\tool, benchmarks, and results are available at \url{www.dsverifier.org}}.

%-----------------------------------
\subsection{Experimental Setup}
\label{experimental-setup}
%-----------------------------------

For all evaluated benchmarks, we generated ``.out'' files that contain 
the counterexamples and verification results ({\it i.e.}, \texttt{successful} and \texttt{failed}) 
produced by DSVerifier~\cite{dsverifier} allied to two verifiers: CBMC~\cite{cbmc} and ESBMC~\cite{esbmc}. 
In addition, the signal input range lies between $-1$ and $1$, that is, the sensor (gyroscope) output bound 
in normal conditions. Using this configuration, inputs employed during the limit-cycle or overflow 
verification is limited between $-1$ and $1$. 
All experiments with \tool v$1$.$1$.$0$  were conducted on an otherwise idle Intel Core i$7-2600$ 
$3.40$ GHz processor, with $24$ GB of RAM, running Ubuntu $64$-bits.
Additionally, the time presented in Table~\ref{table-results} is related to the average of $20$ executions 
for each benchmark; the measuring unit is always in seconds based on the CPU time computed in MATLAB; 
we did not restrict the memory consumption for the experiments.

%%-----------------------------------
%\subsection{Experimental Objectives}
%\label{experimental-objectives}
%%-----------------------------------
%
%\tool was employed to verify the soundness and the reliability of the verification results generated by the DSVerifier tool. 
%Our experimental evaluation aims to answer two research questions:
%%
%\begin{enumerate}
%
%\item[RQ1] \textbf{(performance)} do the executable test cases take considerably less effort
%than verification, {\it i.e.}, it produces a considerably smaller state space?
%
%\item[RQ2] \textbf{(sanity check)} are the counterexamples sound
%and can their reproducibility be confirmed outside of the employed verifier?
%
%\end{enumerate}

%-----------------------------------
\subsection{Experimental Results}
\label{experimental-results}
%-----------------------------------

DSVerifier produced $24$ counterexamples for overflow, 
$27$ for limit-cycle, $54$ for minimum-phase, and $54$ for stability ($159$ counterexamples in total). 
Table~\ref{table-results} shows the \tool results for the quadrotor attitude system digital controllers, 
where ``Property'' describes the property that is evaluated by \tool, ``CE Reproducible'' indicates 
the number of counterexamples that are successfully reproduced, 
``CE Irreproducible'' indicates the number of counterexamples that are not reproducible, 
and ``Time'' provides the runtime in seconds for all simulations on each property. 

\begin{table}[h]
\centering\scriptsize
\caption{Results for the Quadrotor Attitude System.}
\begin{tabular}{r|c|c|c}
Property & CE Reproducible & CE Irreproducible & Time \\
\hline                               
Overflow & 24 & 0 &  0.190\,s\\
Limit Cycle & 26 & 1 & 0.483\,s\\
Minimum-Phase & 54 & 0 & 0.012\,s\\
Stability & 54 & 0 & ~0.188\,s
\label{table-results}
\end{tabular}
\end{table}

Note that the automated validation of all counterexamples took less than $1$ second. 
We consider these times short enough to be of practical use to engineers. 
The present results also show that all counterexamples (except one) generated 
by the underlying verifier, considering FWL effects and different realizations forms 
({\it i.e.}, DFI, DFII and TDFII), are sound and reliable since \tool is able to simulate 
the underlying digital system in MATLAB, and then reproduce the respective counterexample. 
However, for the limit cycle property, there is one counterexample that was not reproduced in \tool; 
DSVerifier did not take into account overflow in intermediate operations to compute the system's output 
using the DFII realization form; this bug was confirmed and fixed by the DSVerifier's developer 
via a github commit.\footnote{\url{https://github.com/ssvlab/dsverifier/commit/88e857bdbc74a7ce3c74d327e2a1e7a246fa48cc}}

\section{Conclusions}
\label{Conclusion}
%-----------------------------------

\tool can validate counterexamples produced for the quadrotor attitude 
system digital controllers, taking into account implementation aspects 
(fixed-point arithmetic and realization), overflow-mode (saturate or wrap-around), 
and rounding-mode (floor and round), to simulate a given digital system with its 
respective counterexample in MATLAB. Currently, \tool is able to perform counterexample 
reproducibility for stability, minimum-phase, limit-cycle, and overflow occurrences. 
Given the current literature in verification, there is no other automated MATLAB toolbox 
to reproduce digital system counterexamples generated by verifiers, and to show the reason 
by which a counterexample cannot be reproduced to validate and endorse the employed 
verification and most importantly to avoid false alarms. 
%
%In addition, \tool is able to reproduce violations for a quadrotor unmanned aerial vehicle (UAV), and our experiments showed that all violations detected for stability, minimum-phase, overflow, and limit-cycle are reproducible, which shows that DSVerifier~\cite{dsverifier} is sound and reliable for those specific properties. 
%
As future work, we expect to expand the field of digital system verification 
with the availability of other verifiers so that \tool can be applied to establish trust in 
untrusted verification results for highly-complex systems. 

%-----------------------------------

%--------------------------------------
\section*{Appendix: DSValidator Demonstration}
%--------------------------------------

%--------------------------------------
\subsection*{Preconditions and Prerequisites}
%--------------------------------------

In order to execute \tool in MATLAB,\footnote{All setup files can be found at~\url{ http://dsverifier.org/.}} user must install the \tool tool in the Linux version of MATLAB 2016a. Each counterexample must be stored as a ``.out'' file extension. If the user has more than one digital system to be validated, a directory with all ``.out'' files must be created.

\noindent \textbf{Tool required:} MATLAB (version: 2016a).

\noindent \textbf{Operating System:} Linux (recommended: Ubuntu 16.04 LTS.)

%--------------------------------------
\subsection*{Installing DSValidator:}
%--------------------------------------

In order to install \tool, user must download the \tool installation file from the DSVerifier web page (\url{http://dsverifier.org/wp-content/uploads/2017/06/dsvalidator-matlab-toolbox-v1.1.0.tar.gz}). After that, the following steps must be executed:

\begin{enumerate}
\item{Open MATLAB and select the folder where you have extracted the \tool tar.gz file.}
\item{Execute the file with the ``$.mltbx$'' extension (or double-click on it); a pop-up screen to install \tool will be shown.}
\item{Click on the ``$install$'' button. After installing \tool, another pop-up screen will be shown to indicate that \tool has been successfully installed in MATLAB.}
\end{enumerate}

User should visualize \tool installed in the $toolbox~menu$ located at the MATLAB toolbar.

%--------------------------------------
\subsection*{Installing DSVerifier:}
%--------------------------------------

In order to install DSVerifier, user must download and extract the DSVerifier installation file from \url{http://dsverifier.org/wp-content/uploads/2017/03/dsverifier-v2.0.3-esbmc-v4.0-cbmc-5.6.tar.gz}. 

\begin{enumerate}
\item{Before using DSVerifier, an environment variable called DSVERIFIER\_HOME must be configured. The user should add it to the .bashrc file as follows:}

\vspace{2 mm}
\noindent  export DSVERIFIER\_HOME='path to dsverifier folder'
\vspace{2 mm}

\item{After setting the environment variable, the following command-line should be used to update the respective variables:}

\vspace{2 mm}
\noindent  {\tt source .bashrc}
\vspace{2 mm}

\item{User should download the (desired) version of CBMC or ESBMC executables for DSVerifier. This package contains CBMC v5.6 and ESBMC v4.0, which should be added to}

\vspace{2 mm}
\noindent  DSVERIFIER\_HOME/model-checker
\vspace{2 mm}

\item{User needs to install the Eigen library (e.g., eigen3, eigen3-static, and eigen3-devel depending on the distribution) and GCC version 5.4 (or higher) if he/she wants to compile the DSVerifier source code for another operating system.}
\end{enumerate}

%--------------------------------------
\subsection*{Preparing the ``.out'' file with the counterexample}
%--------------------------------------

In DSVerifier, users must provide a specification that contains the digital-system numerator and denominator coefficients; the implementation specification that contains the number of bits in the integer and fractional parts; and the input range. For delta realizations, user must also provide the delta operator. In the command-line version, DSVerifier can be invoked as:

\vspace{2 mm}
\noindent \texttt{dsverifier} \emph{\tt <file>} {\tt --realization \emph{\tt <i>} --property \emph{\tt <j>} \\ --x-size \emph{\tt <k>} --BMC \emph{\tt <bmc>} --show-ce-data > file.out}
\vspace{2 mm}

\noindent where \emph{\tt <file>} is the digital-system specification file, \emph{\tt <i>} is the chosen realization form, \emph{\tt <j>} is the property to be verified, \emph{\tt <k>} is the verification bound, {\it i.e.}, the number of times that the digital system is unwound, while the parameter \emph{\tt <bmc>} indicates the BMC tool to be used (CBMC or ESBMC).

For an ANSI-C specification file with many controllers (or filters), additional macro parameters must be provided (\texttt{DDS$\_$ID} and  \texttt{DIMPLEMENTATION$\_$ID}):

\vspace{2 mm}
\noindent \texttt{dsverifier} \emph{\tt <file>} {\tt --realization \emph{\tt <i>} --property \emph{\tt <j>} \\ --x-size \emph{\tt <k>} \hspace{4 mm} --BMC \emph{\tt <bmc>} -DDS$\_$ID=\emph{\tt <l>} \\ -DIMPLEMENTATION$\_$ID=\emph{\tt <m>} --show-ce-data \hspace{4 mm} > file.out}
\vspace{2 mm}

\noindent where \emph{\tt <file>} is the digital-system specification file, \emph{\tt <i>} is the chosen realization form, \emph{\tt <j>} is the property to be verified, \emph{\tt <k>} is the verification bound, \emph{\tt <bmc>} is the BMC tool to be used (CBMC or ESBMC), \emph{\tt <l>} is the digital controller $id$ in the specification file, and \emph{\tt <m>} represents the digital controller implementation.

Currently, $3$ realization forms are supported for digital controllers: direct form I (DFI), direct form II (DFII), and transposed direct form II (TDFII). Additionally, we support delta realizations to check stability and minimum-phase, {\it i.e.}, delta direct form I (DDFI), delta direct form II (DDFII), and transposed delta direct form II (TDDFII).
We also support $4$ different properties to validate the counterexample: overflow, limit cycle, stability, and minimum-phase. 
Fig.~\ref{dsverifer-file} shows an ANSI-C specification file with $1$ controller and $2$ implementations:
\begin{figure}[ht]
\begin{lstlisting}[xleftmargin=.025\textwidth,frame=single,]
#include <dsverifier.h>
#if DS_ID == 1
	digital_system ds = {
		.b = { 1.5, -0.5 },
		.b_size = 2,
		.a = { 1.0, 0.0 },
		.a_size = 2,
		.sample_time = 0.02
	};	
	#define	IMPLEMENTATION_COUNT 2
	#if IMPLEMENTATION_ID == 1
		implementation impl = {
			.int_bits = 2,
			.frac_bits = 14,
			.max = 1.0,
			.min = -1.0,
		};
	#endif	
	#if IMPLEMENTATION_ID == 2
		implementation impl = {
			.int_bits = 4,
			.frac_bits = 12,
			.max = 1.0,
			.min = -1.0,
		};		
	#endif
#endif
\end{lstlisting}
\caption{A digital controller specification file used by DSVerifier.}
\label{dsverifer-file}
\end{figure}

Using the specification shown in Fig.~\ref{dsverifer-file} as an \texttt{input.c} file, DSVerifier can be invoked to check a specific property. The parameter \texttt{--show-ce-data} must be appended to DSVerifier to ensure that the counterexample is provided in a ``.out'' file. In our particular example, the goal is to check for an overflow violation, that considers the digital controller for DFI and DFII realizations, and $2$ different numerical formats (\texttt{<2,14>} and \texttt{<4,14>}). DSVerifer can be invoked as:

\vspace{2 mm}
\noindent \texttt{dsverifier} \emph{\tt input.c} {\tt --realization DFI --property OVERFLOW --x-size 10 --BMC CBMC -DDS$\_$ID=1 -DIMPLEMENTATION$\_$ID=1 --show-ce-data \\ > ds1$\_$impl1$\_$DFI.out}
\vspace{0.5 mm}

To verify the same digital controller, but with the second implementation (\texttt{<4,14>}) and DFI realization:

\vspace{2 mm}
\noindent \texttt{dsverifier} \emph{\tt input.c} {\tt --realization DFI --property OVERFLOW --x-size 10 --BMC CBMC -DDS$\_$ID=1 -DIMPLEMENTATION$\_$ID=2 --show-ce-data \\ > ds1$\_$impl2$\_$DFI.out}
\vspace{0.5 mm}

Using DFII realization to check for an overflow violation in the first implementation \texttt{<2,14>}:

\vspace{2 mm}
\noindent \texttt{dsverifier} \emph{\tt input.c} {\tt --realization DFII --property OVERFLOW --x-size 10 --BMC CBMC -DDS$\_$ID=1 -DIMPLEMENTATION$\_$ID=1 --show-ce-data \\ > ds1$\_$impl1$\_$DFII.out}
\vspace{0.5 mm}

Last, for the second implementation (\texttt{<4,12>}) using DFII realization:

\vspace{2 mm}
\noindent \texttt{dsverifier} \emph{\tt input.c} {\tt --realization DFII --property OVERFLOW --x-size 10 --BMC CBMC -DDS$\_$ID=1 -DIMPLEMENTATION$\_$ID=2 --show-ce-data \\ > ds1$\_$impl2$\_$DFII.out}
\vspace{0.5 mm}

Now, we have $4$ output files that are generated with the respective counterexample for the overflow property, as follows:

\vspace{2 mm}
\noindent {\tt ds1$\_$impl1$\_$DFI.out}, {\tt ds1$\_$impl2$\_$DFI.out}, {\tt ds1$\_$impl1$\_$DFII.out}, {\tt ds1$\_$impl2$\_$DFII.out}
\vspace{0.5 mm}

If we open a ``.out'' file ({\it e.g.}, {\tt ds1$\_$impl1$\_$DFI.out}), then we visualize the counterexample shown in Fig.~\ref{fig-counterexample}.
\begin{figure}[ht]
\begin{lstlisting}[xleftmargin=.025\textwidth,frame=single,]
VERIFICATION FAILED
Counterexample Data:
  Property = OVERFLOW
  Numerator  = { 0.1, -0.09996 }
  Denominator  = { 1.0, -1.0 }
  X Size = 10
  Sample Time = 0.02
  Implementation = <10,6>
  Numerator (fixed-point) = { 384.0, -128.0 }
  Denominator (fixed-point) = { 256.0, 0.0 }
  Realization = DFI
  Dynamic Range = {-1,1}
  Inputs = { 85.328125, -0.0625, 0.0, -128.6875, -215.984375, -256.0, 256.0, -197.359375, 0.0 85.34375 }
  Outputs = { 128.0, -42.765625, 0.03125, -193.03125, -259.640625, -276.0, 512.0, -424.046875, 98.6875, 128.015625 }
\end{lstlisting}
\caption{Counterexample for {\tt ds1$\_$impl1$\_$DFI.out} generated by DSVerifier.}
\label{fig-counterexample}
\end{figure}

\section*{Running DSValidator}

After verifying a given digital-system, one (or more) .out file(s) is (are) generated, which must be saved in a directory, {\it e.g.}, ``/overflow''. In order to check whether a given counterexample is reproducible and reliable, user must invoke \tool in the MATLAB command-line as follows:
\begin{center}
\texttt{validation('/overflow', 'o', '', '', 'ds\_overflow')}
\end{center}

Considering the test cases {\tt ds1$\_$impl1$\_$DFI.out} and 
\\ {\tt ds1$\_$impl1$\_$DFII.out}, the validation output has the following report as shown in Fig.~\ref{fig-report}.
\begin{figure}[ht]
\begin{lstlisting}[xleftmargin=.025\textwidth,frame=single,]
Counterexamples (CE) Validation Report...
CE 1 time: 0.08845 status: reproducible
CE 2 time: 0.010247 status: reproducible
General Report:
Total Counterexamples Reproducible: 2
Total Counterexamples Irreproducible: 0
Total Counterexamples: 2
Total Time Execution: 0.098697
\end{lstlisting}
\caption{Automated Counterexample Validation Report by \tool.}
\label{fig-report}
\end{figure}

A \texttt{ds\_overflow.mat} file is created with $2$ records; the first one is related to {\tt ds1$\_$impl1$\_$DFI.out}, while the second one is related to {\tt ds1$\_$impl1$\_$DFII.out}.

\subsection*{Analyzing the DSValidator Report}

Fig.~\ref{fig-report} shows the report related to the validation process of $2$ test cases; note that in this report, there are $3$ important pieces of information: ``test case $id$'', ``execution time'', and ``validation status''. The ``test case $id$'' represents the test case identification that is validated; ``execution time'' reports in seconds the time to reproduce the test case in MATLAB; ``validation status'' reports \texttt{successful} or \texttt{failed}; \texttt{successful} means that the counterexample produced by DSVerifier is reproducible, while \texttt{failed} means that there is a divergence about the results between MATLAB and DSVerifier.
At the end of the validation report, the total number of \texttt{successful} and \texttt{failed} test cases is provided as well as the total execution time spent during the simulation.

\begin{figure}[ht]
\centering
\includegraphics[width=0.5\textwidth]{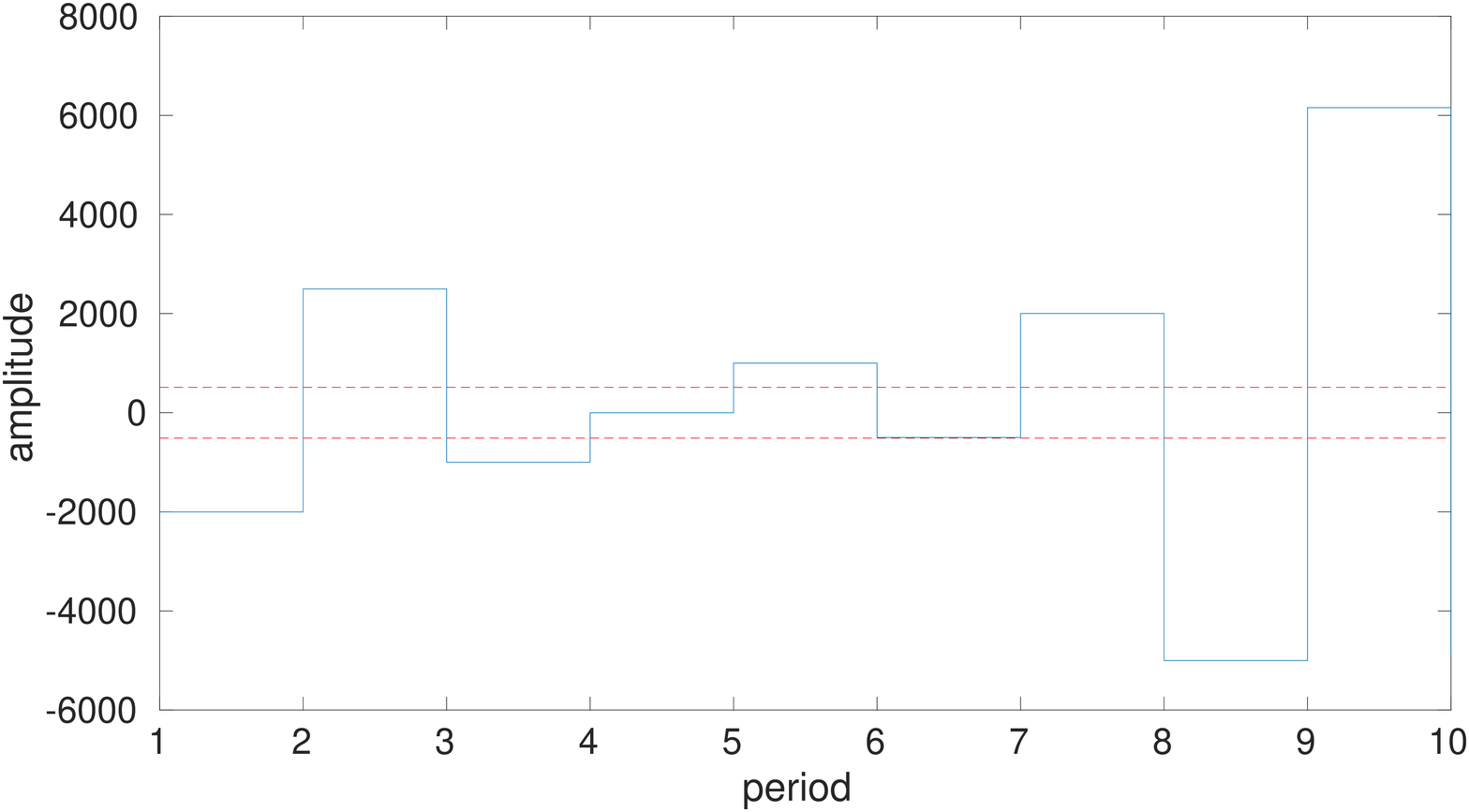}
\caption{Overflow Detection Graphic.}
\label{fig-overflow}
\end{figure}

After the validation process, we can individually check other functions in \tool to inspect a specific step or process. For instance, if we want to inspect the overflow graphic representation (see Fig.~\ref{fig-overflow}), then we can invoke the function \texttt{plot\_overflow(system)} as follows:
\begin{center}
\texttt{plot\_overflow(ds\_overflow(1))}
\end{center}

\begin{figure}[ht]
\centering
\includegraphics[width=0.5\textwidth]{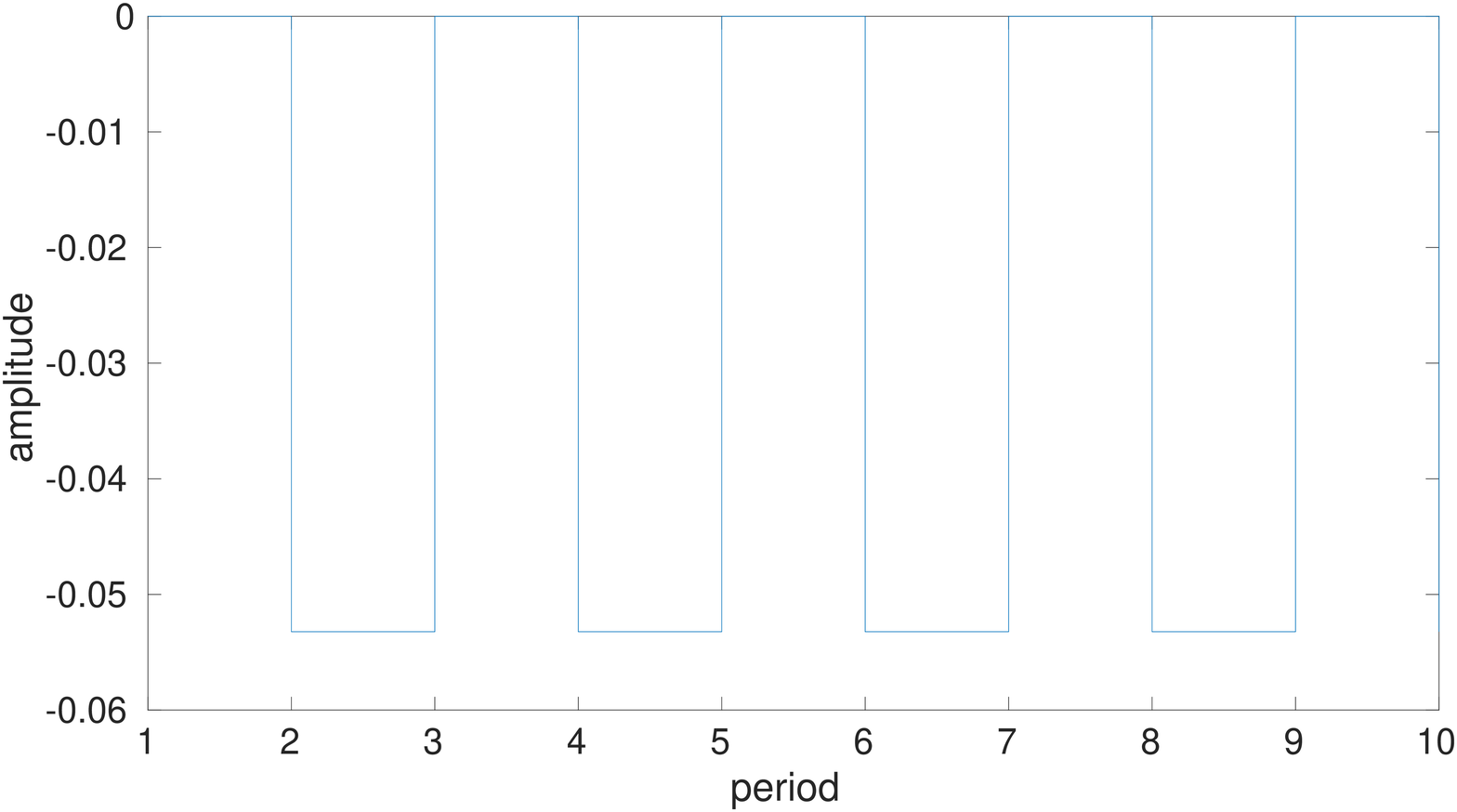}
\caption{Limit Cycle Detection Graphic.}
\label{fig-limit-cycle}
\end{figure}

If we want to visualize the limit cycle graphic representation (see Fig.~\ref{fig-limit-cycle}), then we can invoke the function \texttt{plot\_limit\_cycle(system)} as follows:
\begin{center}
\texttt{plot\_limit\_cycle(ds\_system)}
\end{center}

Other functions from \tool\footnote{All functions are described in \url{http://dsverifier.org/matlab-toolbox/dsvalidator-documentation/}.} can be used to check the counterexamples, {\it e.g.}, there are functions to reproduce the rounding mode, quantization process, and fixed-point operations.
Consider the digital system represented by Eq.~\eqref{eq:eq-pole-zero}:

\begin{equation}
\label{eq:eq-pole-zero}
H(z)=\frac{0.1 + z^{3} - 0.3z^{2} + 0.3{z} - 0.1}{ z^{3} + 1.8z^{2} + 1.14z + 0.272}.
\end{equation}

If we want to check FWL effects on poles and zeros, we must call:

\begin{center}
\texttt{plot\_zero\_pole(ds\_system)}.
\end{center}

\begin{figure}[ht]
    \centering
    \subfloat[Poles and Zeros Graphic without FWL Effects]{
		\includegraphics[width=0.45\textwidth]{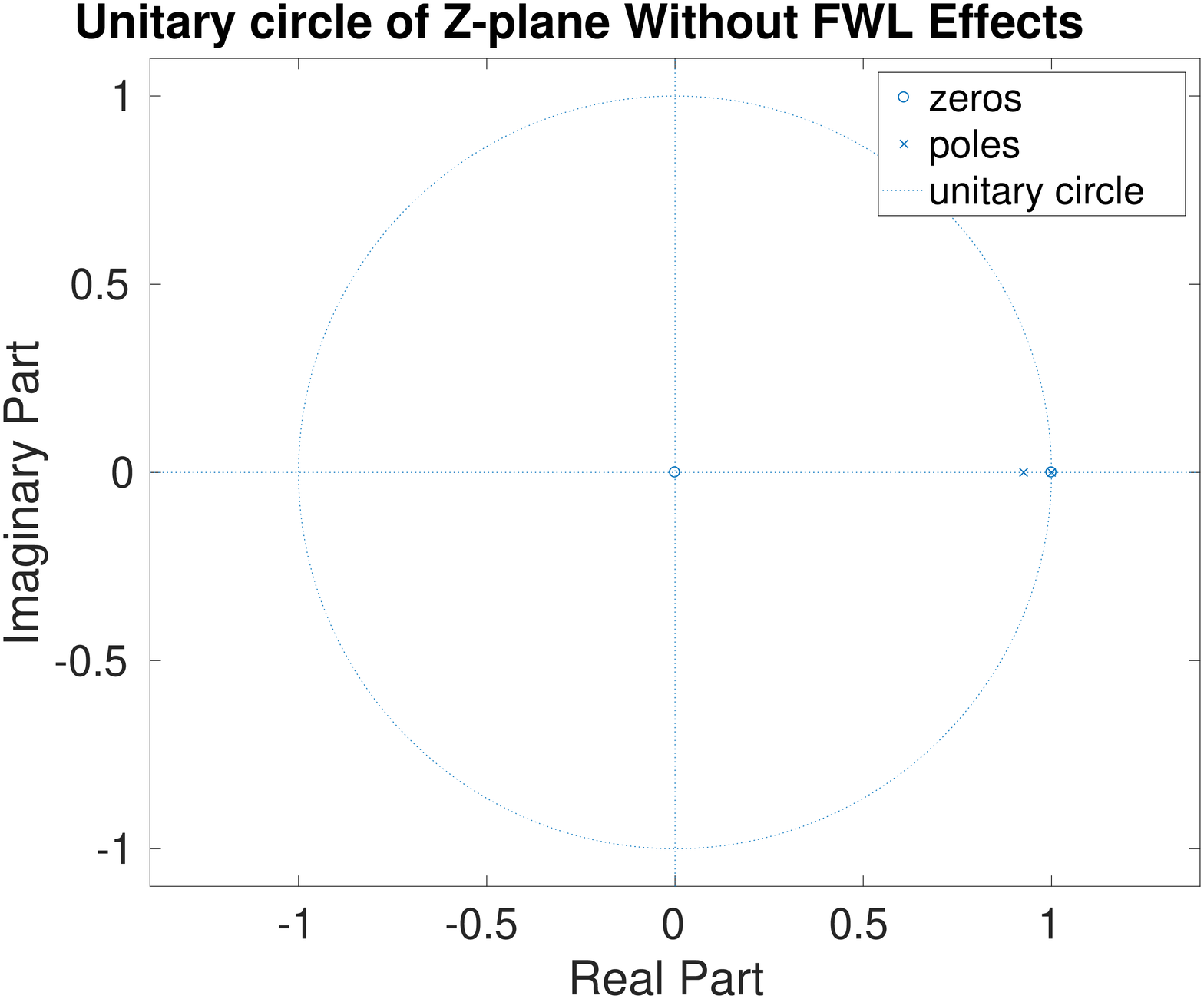}
		\label{pole_zero_subfigure1}}
		\hfil
    \subfloat[Poles and Zeros Graphic with FWL Effects]{
	\includegraphics[width=0.45\textwidth]{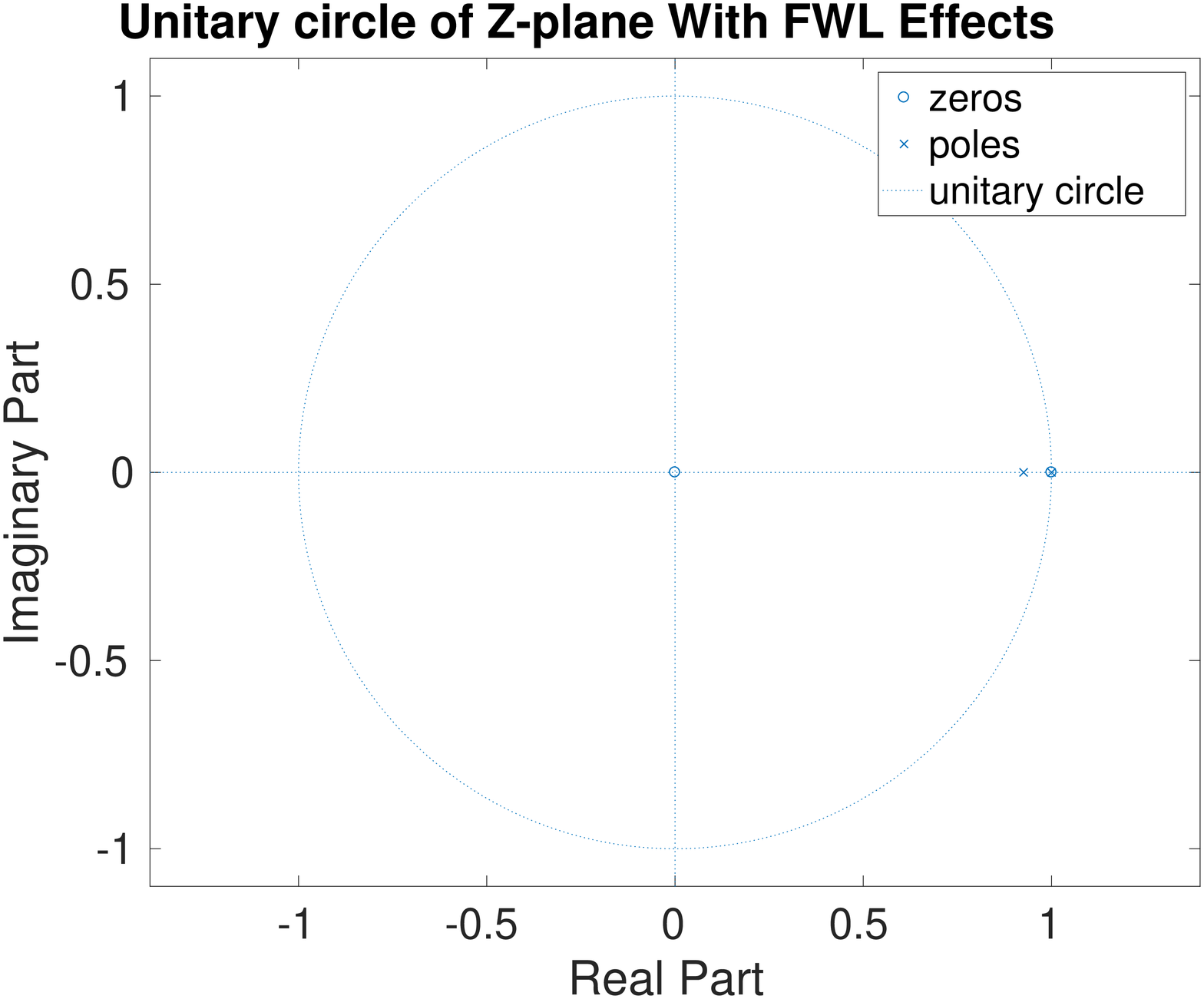}
		\label{pole_zero_subfigure2}}
    \caption{Poles and Zeros Graphic with and without FWL Effects.}
    \label{fig-pole-zero}
 \end{figure}

Fig.~\ref{fig-pole-zero} shows the poles and zeros graphic representation with (and without) FWL effects. Note that when we consider FWL effects in Fig.~\ref{fig-pole-zero}, one zero is out of the $z$-plane unitary circle and thus the controller is a non-minimum-phase system.

We can simulate stability or minimum-phase using the following commands:
\begin{center}
\texttt{simulate\_stability(ds\_system)}
\end{center}
\begin{center}
\texttt{simulate\_minimum\_phase(ds\_system)}
\end{center}

As a result, those functions return \texttt{successful} or \texttt{failed}. If we execute the example for the digital system represented by Eq.~\eqref{eq:eq-pole-zero}, the result for stability is \texttt{successful}, which means that the system is stable. The minimum-phase result is \texttt{failed}, which means that the controller is a non-minimum-phase system.

Another very useful function in \tool is the \texttt{fwl} function that obtains the FWL format of a polynomial, considering implementation aspects. This particular function can be called as:

\begin{center}
\texttt{fwl(poly,l)}
\end{center}
\noindent
where \texttt{poly} represents the polynomial and \texttt{l} represents the fractional bits implementation. 
Considering Eq.~\eqref{eq:eq-poles} as a denominator from a digital system

\begin{equation}
\label{eq:eq-poles}
A(z)= z^{3} + 1.8z^{2} + 1.14z + 0.272,
\end{equation}

\noindent and applying the function \texttt{fwl} in \texttt{A(z)} using $13$ fractional bits, the results is:

\begin{center}
\texttt{>> fwl([1 1.8 1.14 0.272],13)}
\end{center}

\begin{center}
\texttt{>> 1.0000 1.7999 1.1399 0.2720}
\end{center}

\noindent The screencast is available at \url{https://www.youtube.com/playlist?list=PL_Vp1B_NN0gC1-nZtHGHDI2jIN6WvY_6e}

\end{document}